# A Review on Secured Money Transaction with Fingerprint Technique in ATM System

[1] Susmita Mandal

[1] Assistant Professor, Dept. of Information Technology, Marwadi Education Foundation Group of Institutions
Rajkot, Gujarat, India

**Abstract**
In the present day, the requirement of securing electronic cash flow system are increasing significantly. Today's life is so busy that spending a valuable second cost so much. In, such situation if money flow is possible swiftly by just one swipe it would be a great relief. Biometric based authentication can be a new approach to satisfy user needs by replacing password-based authentication. Among all biometric techniques fingerprinting is the oldest and secured methodology practised till date. In the proposed system user can transact money by placing his/her thumb imprint on new proposed ATM card. This new system will smoothen the transaction with security.

## 1. Introduction

Today's world is fast paced we need things to be done swiftly and quick. To achieve the necessity of the mass scientists have invented new machinery to smoothen the work. For the same security has been kept high through PINs (Personnel identification number) and password for operating multiple devices like car, radio ,mobile and ATM machines herein it is a major difficulty facing by customer like usability, memorability and security. Password makes difficulty to to remember, some people write their passwords on piece of paper or notebook or keep as remember password on browsers while surfing which is vulnerable. Users are allowed to choose their own passwords which would be easy to remember but also can be guessed by brute force attackers.

An automatic teller machine or ATM allows a bank customer to conduct their banking transactions from almost every other ATM machine in the world like, deposits ,transfers, balance enquires and withdrawal. Crimes at ATM's has become a nationwide issue that faces not only customers but also bank operators[1]. Security measures at banks can play a critical, contributory role in preventing attacks on customers.

Authentication methods for ATM cards have little changed since their introduction in the 1960's. typically, the authentication design involves a trusted hardware device (ATM card or token).The card holder's Personal identification number(PIN) is usually the one means to verify the identity of the user. However ,due to the limitations of such design ,an intruder in possession of user's device can discover the user's PIN with brute force attack. For instance, in a typical four digit Pin, one in every 10,000 users will have the same number.

Despite many security measures, cases of ATM crime continue to occur globally. Incidents have been reported in Asia-Pacific, the Americans, Africa, Russia and Middle East[2].

In the upcoming sections detailed description about types of ATM Threats , ATM architecture, Modularized ATM card with fingerprint technique, UML designs for the proposed system. Finally, conclusion of the paper.

## 2. ATM Threats

ATM threats are divided into 3 categories: card and currency fraud, logical attacks and physical attacks.

2.1 Card and Currency Frauds

Card and currency fraud involves both direct attacks to steal cash from the ATM and indirect attacks to steal a consumer's identity (in the form of consumer card data and PIN theft). The intent of indirect attacks is to fraudulently use the consumer data to create counterfeit cards and obtain money from the consumer's account through fraudulent redemption.

2.1.1 Skimming

ATM card skimming is the most prevalent and well known attack against ATMs. Card skimmers are devices used by perpetrators to capture cardholder data from the magnetic stripe on the back of an ATM card. It resembles like a hand-held credit card scanner—are often installed inside or over top of an ATM's factory-installed card reader[3].

2.1.2 Card Trapping/Fishing

The purpose of this type of attack is to steal the card and use it at a later time to make fraudulent withdrawals from the consumers' compromised accounts . This involves placing a device over or into the ATM card reader slot .In, this case a card is physically captured by trapper.

2.2 Logical/data Attacks

The most difficult attacks to detect, logical attacks target an ATM's software, operating system and communications systems.

### 2.2.1 Pin Cracking

It attacks targets the translate function in switches- an abuse functions that are used to allow customers to select their PINs online. A banker inside could use an existing Hardware Security Module(HSM) to reveal the encrypted PIN codes.

### 2.2.2 Malware and Hacking

With any computer, the purpose of installing malicious software(malware) is to violate the confidentiality, integrity and authenticity of data on the computer system. Attackers use sophisticated programming techniques to break into websites which reside on a financial institution's network.

### 2.3 Physical Attack

ATM physical attacks are attempted on the safe inside the ATM, through mechanical or thermal means with the intention of breaking the safe to collect the cash inside. The methods of attacks used to try to gain access to the safe include:

- Cutting/grinding – usually with power saws and grinders
- Drilling – usually with power drills
- Prying – with pry bars, wedges, and crowbars
- Pulling – after the safe door has been cut with a saw or torch, one end of a chain or cable is connected to the door and the other end to a vehicle to pull off the door.
- Torch or other burning device such as a thermal lance
- Explosives such as gas, dynamite, homemade bombs, or even gasoline.

## 3. Related Work

This section includes the proposed design along with new devices required for implementing the model.

### 3.1 ATM Card with Integrated Fingerprint Sensor

The improved ATM card preferably comprises a CPU, memory, and a fingerprint reader including a sensing surface. Preferably, the sensing surface is located along a surface of the smart card so that a user's thumb is naturally positioned over the sensing surface when the card is inserted into a read unit of the card reader. When an individual inserts the card into a read unit, the ATM card creates an electrical representation of the individual's fingerprint and compares the acquired representation to a stored fingerprint representation in the card's memory. If the acquired representation matches the stored representation, the card is enabled, and the user is given access to information and/or services that require cooperation of the smart card[6].

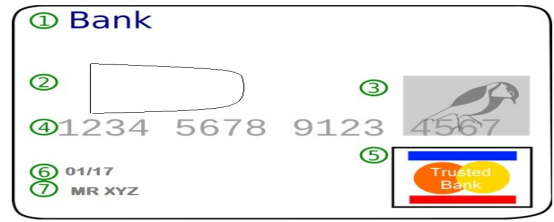

Fig.1 Modularized ATM card with fingerprint sensor

The front side of card consists of following things:-

1. Issuing bank logo
2. Fingerprint sensor
3. Hologram
4. Card number
5. Card brand logo
6. Expiration date
7. Card holder name

### 3.2 Modularized Card Reader

The new ATM card reader has an input terminal to read the card inserted. The card reader verifies the fingerprint generated while pressing user thumb on the card while insertion. As, the card already has a memory unit which stores the user's actual fingerprint during opening an account in bank. The stored thumb print is matched with the new impressed thumbprint.

Once acknowledged the user is allowed for processing further transaction by selecting his/her saving account from the server.

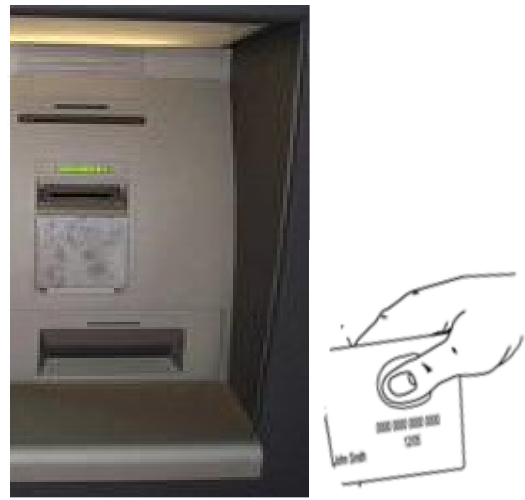

Fig.2 Modularized fingerprint verifying card reader

### 3.3 ATM Architecture

In general ATM architecture[7] consist of a smart card reader and a Keypad to enter the 4 digit PIN (personal identification number) which allows a user to access his/her account if verified.

In the proposed architecture a user in inserting a card with thumb impression on it in the fingerprint card reader slot which generate an electric pulse to verify the stored fingerprint with the original. Once, verified the user is allowed to perform transaction in his/her account as usual.

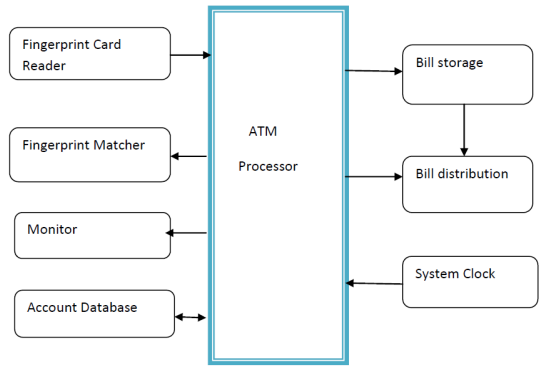

Fig.3 Proposed ATM Architecture

### 3.4 Fingerprint Verification Process

The verification process is explained using use case diagram from customer perspective [4].

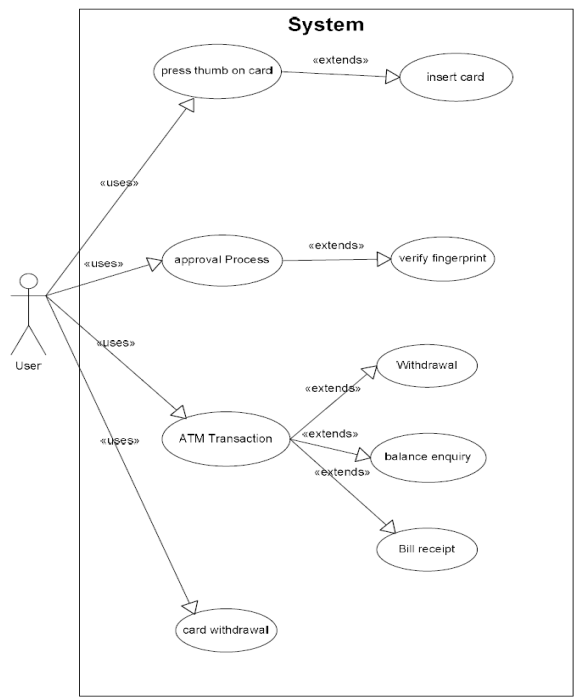

Fig.4 UML diagram for amount transaction

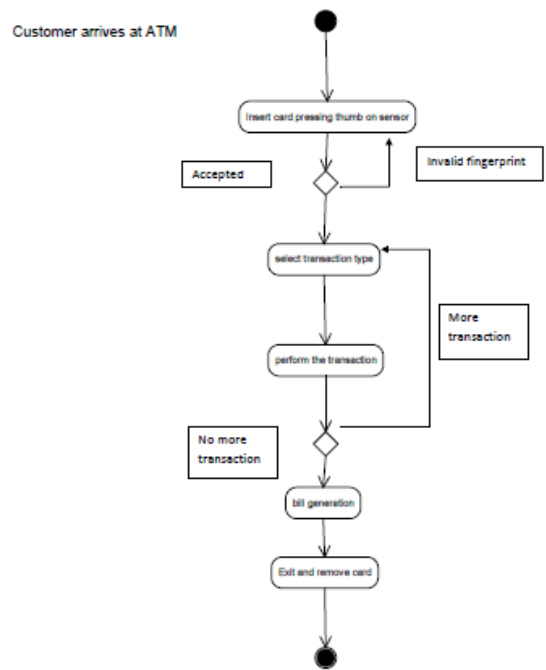

Fig.5 Activity diagram of transaction and verification process

### 3.5 Advantages/Disadvantages of the Model

Like, any other technology biometric has its own advantages and disadvantages. It is an upcoming technology, bank can apply for effective security measures with easy transaction [5].

1. It provides strong authentication.
2. It works accurately.
3. It easy to operate.
4. Flexibility to any time operating system.
5. Various type of hidden expenses can be avoided.
6. No need to remember PIN number.
7. Fast enough speed.

Certain limitations noted down are as following.

1. It depends on user acceptability.
2. System cost are biggest technical problem.
3. Fingerprint worn with hard labour work or age.
4. Biometric ATM's are expensive security solutions.

## 4. Conclusion

This biometric ATM system is highly secure because it works information contained within body parts. Biometrics is uniquely bound to individuals and may offer organizations a stronger method of authentication and verification. Biometric ATM is very useful also very difficult to implement . But for security purpose or control the criminal offences it is very important and helpful method .

In future many other less expensive methods can be experienced using biometric technique to solve major ATM frauds occurring in today's world.

## References

[1] Richard . B. and Alemayehu, M.(2006) Developing E-banking Capabilities in a Ghanaian Bank:Preliminary Lessons.Journal of Internet banking and commerce August 2006,vol.11,no.2. available online (http://www.arraydev.com/commerce/jibc/) Accessed on 24/11/2009.

[2] ATM market place.(2009a)." ATM scam nets Melbourne thieves $500,000',"retrieved december 2,2009, from http://www.atmmarketplace.com/article.php?id=10808

[3] http://www.diebold.com/atmsecurity/files/DBD_ATMFraud_WP.pdf

[4] Enhanced atm security system using biometrics IJcSi vol.9,issue 5,no 3,September 2012 (www.IJCSI.org)

[5] Use of biometrics to tackle ATM fraud 2010 international conerence on business and economics research vol.1(2011)

[6] Patent *US6325285 B1* - Smart card *with integrated* fingerprint inventor Paul J. Baratelli and At&T Corp.

[7] The Formal Design Model of an Automatic Teller Machine (ATM) Yingxu Wang, *University of Calgary, Canada*

**Author** Susmita Mandal M.Tech in Information Security and Computer Forensic, B.Tech in Information Technology.Worked as Guest Lecturer at Dr.B.R.Ambedkar Govt.Polytechnic ,Port Blair,India .Currently resuming the post of Assistant Professor at Marwadi Education Foundation Group of Institution, Rajkot,India. Published a Book on A Key Management Solution for Reinforcing Compliance with HIPAA Privacy/Security Regulations in December 2012 by Lambert Academic Publishing .Current research interest are cloud computing, Information retrieval, network security.